\documentclass[12pt]{article}
\usepackage{graphicx}
\usepackage{psfrag}
\renewcommand{\nu}{v}
\newcommand{\spacing}[1]{\renewcommand{\baselinestretch}{#1}\large\normalsize}  
\spacing{1.0} % double spacing
\begin{document}
\author{Daniel Duque, Xiao-jun Li%
\footnote{Current address: The Institute for Systems Biology,
4225 Roosevelt Way NE, Suite 200, Seattle, Washington 98105-6099, USA.},
Kirill Katsov, and M. Schick\\
        Department of Physics, Box 351560 \\
        University of Washington, Seattle 98195-1560}
\title{Molecular theory  of 
hydrophobic mismatch between lipids and peptides}
%\vskip .4truein
%Running Title: Molecular theory of hydrophobic mismatch}
\date{\today}

\maketitle

\begin{abstract}
Effects of the mismatch between the hydrophobic length, $d$, of 
transmembrane alpha helices of 
integral proteins and the hydrophobic thickness, $D_h$, of the membranes they
span are studied theoretically utilizing a microscopic model of
lipids. In particular, we examine the dependence of the period of a
lamellar phase on the hydrophobic length and volume fraction 
of a rigid, integral, peptide. 
We find that the period decreases when a short
peptide, such that $d<D_h$, is inserted.
More surprising, we find that the period increases
when a long peptide, such that $d>D_h$, is inserted. The effect is due
to the replacement of extensible lipid tails by rigid peptide. 
As the peptide length is increased, the lamellar 
period continues to increase, but at a slower rate, and can eventually
decrease. The amount of peptide which fails to incorporate 
and span the membrane increases
with the magnitude of the hydrophobic mismatch $|d-D_h|$.
We explicate these behaviors which are all in accord
with experiment. Predictions are made for the dependence of the tilt of
a single trans-membrane alpha helix on hydrophobic mismatch and helix density.
\end{abstract}

\section{Introduction} 

The interaction of integral proteins with the lipids in which they are
embedded is of great importance for membrane function (Dumas et al.,
1999). One principal governing this interaction is that the length of
the hydrophobic segment of protein should closely match that of the
membrane which it spans (Bloom and Mouritsen, 1984). Among the evidence
that this hydrophobic matching is used in membrane organization is that
the various membranes between the Golgi and the plasma membrane have
different thicknesses. Proteins can be routed through the secretory
pathway by increasing their hydrophobic thickness via mutagenesis and
passing from one membrane to another more closely matching their new
length (Pelham and Munro 93). The difference between the hydrophobic
length of protein and membrane, denoted hydrophobic mismatch, affects
{\em inter alia}, lateral segregation of proteins in membranes (Marsh
1995; Lehtonen and Kinnunen 1997), the lipid melting transition (Piknova
et al., 1993), and protein activity (Johannsson et al. 1981: Froud et
al. 1986).
Hydrophobic mismatch also affects the way in which
the stabilty and the inclination 
of transmembrane helices change as functions of their hydrophobic length. 
Such information is very important in predicting transmembrane domains 
from potential protein sequences (Rost et al. 1995; Edelman 1993), 
a topic becoming increasely important in Biology with the completion of 
the Human Genome Project. 

In an effort to elucidate such effects on a molecular level, Killian and
co-workers (de Planque et al. 1998) investigated the effect of a series
of hydrophobic peptides on the mean thickness of phosphatidylcholine
membranes with different tail lengths. The peptides consisted of a
sequence, whose length could be adjusted, of alternating leucine and
alanine flanked on both sides by two tryptophans. The latter prefer to
reside just below the lipid head groups, and therefore serve as anchors
for the peptide (Killian et al., 1996). The N- and C- terminii
were blocked, {\em e.g} FmAW$_2$(LA)$_n$W$_2$Etn.
The results of this study which are of the most interest
to us are as follows:

\begin{itemize}
\item peptides whose hydrophobic thickness, $d$, is smaller than that of the
hydrophobic thickness, 
$D_h$,
of the bilayer cause a reduction in the bilayer
thickness.

\item peptides whose hydrophobic thickness is larger than that of
the bilayer 
cause an
increase in the bilayer thickness. As the peptides are made longer, the
membranes continue to thicken, but the increase in the membrane thickness
is always less than the peptide increment.

\item an increase in the magnitude of mismatch, $|d-D_h|$, 
whether due to peptides
being too long or too short for the membrane, results in an increase in
the fraction of peptides which fail to incorporate into the
membrane. This effect has also been observed by Ren et al. (Ren et
al. 1997, 1999).
\end{itemize}

The first observation is easily understood on simple energetic
arguments, but the second is not. One might have expected the longer
peptide simply to insert at an angle such that the membrane thickness is
undisturbed (Killian, 1998). However the thickness {\em is} disturbed, and grows with
increased peptide length, but does not track that increase
identically. To understand this puzzling behavior and to isolate the
various factors which bring about the net result of hydrophobic mismatch
is the primary purpose of this paper.

Many theoretical approaches have been applied to the effects of
mismatch. Almost all of them are phenomenological, based on treating the
membrane as an elastic sheet. Some have, however, included in these
phenomenological descriptions some properties of the lipid tails, such
as the ability of the tails to tilt to accommodate the perturbation of
inserted proteins (May and Ben-Shaul, 1999). These approaches, which
have recently been reviewed (May 2000), are valuable in clarifying
several aspects of the general problem of membrane impurities, and have
even been applied (May and Ben-Shaul, 1999) to the lamellar- inverted
hexagonal transition induced by a sufficient concentration of peptide
(van der Wei et al., 2000). Nevertheless, they lack a direct link to the
molecular details of the system.  A major theoretical advance was the
work of Fattal and Ben-Shaul (Fattal and Ben-Shaul, 1993) who provided a
molecular theory for the behavior of the lipid chains of the
membrane. The peptides, with their hydrophobic length, were treated as
providing a boundary condition on the configuration of the lipid
chains. This molecular modelling was combined with phenomenological free
energy contributions describing lipid headgroup repulsion and membrane
solvent surface tension.  In this paper we eschew phenomenological
description and present a molecular theory which, from straight forward
statistical mechanics, yields the free energy of the entire system,
lipid and peptide.  We utilize a molecular lipid model employed earlier
(Li and Schick, 2000) and treat the peptide, which traverses the membrane as
an alpha helix, as a rigid rod.  We consider a lamellar phase formed by the
lipid, and investigate the effect on the period of this phase due to the
addition of peptide in small concentrations. As a consequence of our
calculations, we are able to reproduce all of the above results,
elucidate the reasons for the increase in membrane thickness when
penetrated by long peptides, and delineate several different effect of
hydrophobic mismatch.

\section{The Model and its Self Consistent Field Solution}

We consider an anhydrous system of volume $V$ consisting of lipids and
peptides whose numbers, $N_\mathrm{l}$ and $N_\mathrm{p}$,
are controlled by the fugacities $z_\mathrm{l}$
and $z_\mathrm{p}$, respectively.  The lipids consist of a headgroup of
volume $\nu_\mathrm{h}$, and two equal-length tails each consisting of
$N$ segments of volume $\nu_\mathrm{t}$. Each lipid tail is
characterized by a radius of gyration $R_\mathrm{g}=(Na^2/6)^{1/2}$,
with $a$ the statistical segment length. The peptide consists of a
rigid, hydrophobic, core of $L$ segments each of length $b$ and volume
$\nu_\mathrm{c}$ and a terminal group at each end of volume
$\nu_\mathrm{e}$. We choose these end segments to be hydrophilic so that
the peptide indeed models a hydrophobic segment within an otherwise
hydrophilic region. The hydrophobic length of the peptide is $d=Lb$.

There are four local densities which specify the state of the system. We
measure them all with respect to the convenient density
$\nu_\mathrm{h}^{-1}$. They are the number density of the lipid
headgroup, $\nu_\mathrm{h}^{-1}\Phi_\mathrm{h}(\mathbf{r})$, and of the
lipid tail segments, $\nu_\mathrm{h}^{-1}\Phi_\mathrm{t}(\mathbf{r})$,
the number density of the peptide core segments,
$\nu_\mathrm{h}^{-1}\Phi_\mathrm{c}(\mathbf{r})$, and the peptide end
groups, $\nu_\mathrm{h}^{-1}\Phi_\mathrm{e}(\mathbf{r})$.  Note that all
number densities, $\Phi(\mathbf{r})$, are defined to be dimensionless.

We consider repulsive contact interactions between these four
elements. In the simplest case, the strengths of the 
repulsive interactions between the two hydrophilic and the two hydrophobic elements
are the same, $kT\nu_\mathrm{h}\chi$, with $k$ Boltzmann's constant, and
$T$ the absolute temperature. We take the total energy of interaction to be
\begin{equation}
\label{energy}
 E[\Phi_\mathrm{h},\Phi_\mathrm{t},\Phi_\mathrm{e},\Phi_\mathrm{c}]=kT\chi\int
{d\mathbf{r}\over \nu_\mathrm{h}}
[\Phi_\mathrm{h}(\mathbf{r})+\gamma_\mathrm{e}\Phi_\mathrm{e}
(\mathbf{r})]
[\gamma_\mathrm{t}\Phi_\mathrm{t}(\mathbf{r})+\gamma_\mathrm{c}
\Phi_\mathrm{c}(\mathbf{r})];
\end{equation} 
where we have introduced the relative volume of the
tails, $\gamma_\mathrm{t}=2N\nu_\mathrm{t}/\nu_\mathrm{h}$, of the
peptide cores, $\gamma_\mathrm{c}=L\nu_\mathrm{c}/\nu_\mathrm{h}$, and
of the peptide end groups
$\gamma_\mathrm{e}=\nu_\mathrm{e}/\nu_\mathrm{h}$. Note that the
interaction energies depend upon the local volume fractions
$\gamma_e\Phi_e$ etc. as opposed to the local number densities $\Phi_e$
etc.  (Williams and Fredrickson, 1992).

In addition to this local repulsion, we include the hard core
interactions between all particles in an approximate way by imposition
of a local incompressibility constraint, \textit{i.e.}, that the sum of
the volume fractions of all components must be unity everywhere:
\begin{eqnarray} 
\label{incomp}
\Delta({\bf r})&\equiv &\Phi_\mathrm{h}(\mathbf{r})+
\gamma_\mathrm{t}\Phi_\mathrm{t}(\mathbf{r})
+\gamma_\mathrm{e}\Phi_\mathrm{e}(\mathbf{r})+
\gamma_\mathrm{c}\Phi_\mathrm{c}(\mathbf{r})-1\nonumber\\
               &=&0. 
\end{eqnarray}
As shown earlier (Li and Schick, 2000), the partition function of the
system can be written in the form in which the four fluctuating
densities, instead of interacting directly with one another, interact
indirectly via four fluctuating fields, here denoted $W_\mathrm{h}$,
$W_\mathrm{t}$, $W_\mathrm{e}$, and $W_\mathrm{c}$.  Self consistent
field theory results when the fluctuating fields and densities are
approximated by those values which extremize the free energy, $\Omega$,
of the system in the presence of these fields. This free energy has the
form
\begin{eqnarray}
\label{free1}
{\nu_\mathrm{h}\over kTV}{\Omega} 
&=&-z_\mathrm{l}{{\cal Q}_\mathrm{l}[W_\mathrm{h},W_\mathrm{t}]\over V}-
z_\mathrm{p}{{\cal Q}_\mathrm{p}[W_\mathrm{e},W_\mathrm{c}]\over V}
 +{\nu_\mathrm{h}\over kTV}E\nonumber \\
&-&\int{d\mathbf{r}\over V}[W_\mathrm{h}(\mathbf{r})
\Phi_\mathrm{h}(\mathbf{r})+
W_\mathrm{t}(\mathbf{r})\Phi_\mathrm{t}(\mathbf{r})
+ W_\mathrm{e}(\mathbf{r})\Phi_\mathrm{e}(\mathbf{r}) 
+W_\mathrm{c}(\mathbf{r})\Phi_\mathrm{c}(\mathbf{r})]\nonumber \\
&-&\int{d\mathbf{r}\over V}\Xi(\mathbf{r})\Delta(\mathbf{r}).
\end{eqnarray}
Here ${\cal Q}_\mathrm{l}[W_\mathrm{h},W_\mathrm{t}]$ is the partition
function of a \emph{single} lipid in external fields $W_\mathrm{h}$, and
$W_\mathrm{t}$, and ${\cal Q}_\mathrm{p}[W_\mathrm{e},W_\mathrm{c}]$ is
the partition function of a \emph{single} peptide in the external fields
$W_\mathrm{e}$ and $W_\mathrm{c}$.  Note that a Lagrange multiplier
$\Xi(\mathbf{r})$ has been introduced to enforce the incompressibility
constraint of Eq. \ref{incomp}.  The functions $W_\mathrm{h}$,
$\Phi_\mathrm{h}$ \textit{etc}. which extremize this free energy
will be denoted by their corresponding lower case letters $w_\mathrm{h}$
and $\phi_\mathrm{h}$. They are obtained from the five equations for the
fields
\begin{eqnarray}
\label{scf1}
w_\mathrm{h}(\mathbf{r})&=&\chi[\gamma_\mathrm{t}
\phi_\mathrm{t}(\mathbf{r})+\gamma_\mathrm{c}\phi_\mathrm{c}(\mathbf{r})]+
\xi(\mathbf{r}),
\\%
\label{scf2}
w_\mathrm{t}(\mathbf{r})/\gamma_\mathrm{t}&=&\chi[\phi_\mathrm{h}
(\mathbf{r})+\gamma_\mathrm{e}\phi_\mathrm{e}(\mathbf{r})]+\xi(\mathbf{r}),
\\%
\label{scf3}
w_\mathrm{e}(\mathbf{r})/\gamma_\mathrm{e}&=&\chi
[\gamma_\mathrm{t}\phi_\mathrm{t}(\mathbf{r})+\gamma_\mathrm{c}
\phi_\mathrm{c}(\mathbf{r})]+\xi(\mathbf{r}),
\\%
\label{scf4}
w_\mathrm{c}(\mathbf{r})/\gamma_\mathrm{c}&=&\chi[\phi_\mathrm{h}(\mathbf{r})+
\gamma_\mathrm{e}\phi_\mathrm{e}(\mathbf{r})]+\xi(\mathbf{r}),
\\%
\label{scf5}
1&=&\phi_\mathrm{h}(\mathbf{r})+\gamma_\mathrm{t}\phi_\mathrm{t}(\mathbf{r})
+\gamma_\mathrm{e}\phi_\mathrm{e}(\mathbf{r})+
\gamma_\mathrm{c}\phi_\mathrm{c}(\mathbf{r}).
\end{eqnarray}        
          
The field $\xi$ can be easily eliminated, while Eqs. (\ref{scf2}) and
(\ref{scf4}) imply $w_\mathrm{t}(\mathbf{r})/\gamma_\mathrm{t}=
w_\mathrm{c}(\mathbf{r})/\gamma_\mathrm{c}$, so that one deals
essentially with three equations. The four densities are all functionals
of the above fields except $\xi$ and, therefore, close the cycle of
self-consistent equations:
\begin{eqnarray}
\label{head}
\phi_\mathrm{h}(\mathbf{r})[w_\mathrm{h},w_\mathrm{t}]&=
&-z_\mathrm{l}{\delta{\cal Q}_\mathrm{l}[w_\mathrm{h},w_\mathrm{t}]\over\delta 
w_\mathrm{h}(\mathbf{r})},\\
\label{tail}
\phi_\mathrm{t}(\mathbf{r})[w_\mathrm{h},w_\mathrm{t}]&=&
-z_\mathrm{l}{\delta{\cal Q}_\mathrm{l}[w_\mathrm{h},w_\mathrm{t}]\over\delta 
w_\mathrm{t}(\mathbf{r})}, \\
\label{endden}
\phi_\mathrm{e}(\mathbf{r})[w_\mathrm{e},w_\mathrm{c}]&=&
-z_\mathrm{p}{\delta{\cal Q}_\mathrm{p}[w_\mathrm{e},w_\mathrm{c}]\over\delta 
w_\mathrm{e}(\mathbf{r})},\\
\label{coreden}
\phi_\mathrm{c}(\mathbf{r})[w_\mathrm{e},w_\mathrm{c}]&=&
-z_\mathrm{p}{\delta{\cal Q}_\mathrm{p}[w_\mathrm{e},w_\mathrm{c}]\over\delta 
w_\mathrm{c}(\mathbf{r})}.
\end{eqnarray}
With the aid of the above equations, the self consistent, or mean field,
free energy $\Omega_\mathrm{mf}$, which is the free energy function of
Eq.  \ref{free1} evaluated at the self consistent field values of the
densities and fields, can be put in the form
\begin{equation}
\label{omega}
-\Omega_{\mathrm{mf}}
(z_\mathrm{l},z_\mathrm{p},T)
={kT\over \nu_\mathrm{h}}\left(z_\mathrm{l}
{\cal Q}_\mathrm{l}[w_\mathrm{h},w_\mathrm{t}]+z_\mathrm{p}{\cal
Q}_\mathrm{p}[w_\mathrm{e},w_\mathrm{c}] +V\xi_0\right) 
+E[\phi_\mathrm{h},\phi_\mathrm{t},\phi_\mathrm{e},\phi_\mathrm{c}]
\end{equation}
where we have defined $V\xi_0\equiv\int\xi({\bf r})d{\bf r}$. 
All the fields, $w$, $\xi_0$, and densities, $\phi$, are functions of the
activities, $z_\mathrm{l}$, $z_\mathrm{p}$ and temperature.
%From this thermodynamic potential, the Helmholtz free energy can be obtained,
%\begin{eqnarray} 
%F(N_\mathrm{l},N_\mathrm{p},T)&=&\Omega+kTN_\mathrm{l}\ln(z_\mathrm{l}
%\lambda^3_\mathrm{l}/v_\mathrm{h})+
%kTN_\mathrm{p}\ln(z_\mathrm{p}\lambda^3_\mathrm{p}/v_h)\nonumber\\
%                              &=&N_\mathrm{l}f_\mathrm{l}+
%N_\mathrm{p}f_\mathrm{p}-E\nonumber\\
%&+&kTN_\mathrm{l}[\ln(N_\mathrm{l}\lambda^3_\mathrm{l}/V)-1-
%(\xi_0/\rho_0\nu_\mathrm{h} )]\nonumber\\
%&+&kTN_\mathrm{p}[\ln(N_\mathrm{p}\lambda^3_\mathrm{p}/V)-1- (\xi_0/\rho_0 \nu_\mathrm{h})],
%\end{eqnarray}
%where $\lambda_\mathrm{l}$ and$\lambda_\mathrm{p}$ are the de Broglie 
%wavelengths of lipid and peptide,
%$\rho_0=(N_\mathrm{l}+N_\mathrm{p})/V$ is the 
%number density of the system, and 
%$f_\mathrm{l}=-kT\ln({\cal Q}_\mathrm{l}/V)$ and
%$f_\mathrm{p}=-kT\ln({\cal Q}_\mathrm{p}/V)$ are the free energies of
%single lipids and peptides in the external fields.
All of the above is a simple extension of previous procedure 
(Li and Schick, 2000).
 
There remains only to specify how
the partition functions of the lipids and of the peptides
are calculated. 
One defines the end-segment distribution function of the lipid 
$q(\mathbf{r},s)$. Because 
the lipid tails are treated as completely flexible, this
function satisfies the modified diffusion equation
\begin{equation}
{\partial q(\mathbf{r},s)\over \partial s}-
2R_\mathrm{g}^2\nabla^2q(\mathbf{
r},s)
+\left[w_\mathrm{h}(\mathbf{r})\delta(s-1/2)+w_\mathrm{t}(\mathbf{
r})\right]q(\mathbf{r},s)=0,
\end{equation}
with initial condition
\begin{equation} 
q(\mathbf{r},0)=1.
\end{equation}
From this function, one obtains the partition functions of the lipids, 
\begin{equation}
{\cal Q}_\mathrm{l}=\int d\mathbf{r}\ q(\mathbf{r},1), 
\end{equation} 
and, from Eqs. (\ref{head}) and (\ref{tail}), the head and tail
densities
\begin{eqnarray}
\label{phih}
\phi_\mathrm{h}(\mathbf{r})&=&z_l\exp\{-w_\mathrm{h}(\mathbf{r})\}
q\left(\mathbf{r},{1\over
2}-\right)
q\left(\mathbf{r},{1\over 2}-\right),\\
\label{phit}
\phi_\mathrm{t}(\mathbf{r})&=&z_\mathrm{l}\int_\mathrm{0}^1ds
\ q(\mathbf{r},s)q(\mathbf{
r},1-s)
\end{eqnarray}
To obtain the partition function of the peptide, one defines its
end-segment distribution function 
$q_\mathrm{p}(\mathbf{r},\hat{\mathbf{n}},s)$, where
$\hat{\mathbf{n}}$ is a unit vector which specifies the orientation of the
peptide. Because the peptide is rigid, and of length $Lb$, 
this function satisfies the
equation  (Wang and Warner, 1986)
\begin{equation}
\label{pepend}
{\partial q_\mathrm{p}(\mathbf{r},\hat{\mathbf{n}},s)\over \partial s}+
L b \hat{\mathbf{n}}\cdot \nabla_\mathbf{r} 
q_\mathrm{p}(\mathbf{r},\hat{\mathbf{n}},s)+
\{w_\mathrm{e}(\mathbf{r})[\delta(s)+\delta(s-1)]
+w_\mathrm{c}(\mathbf{r})\}q_\mathrm{p}(\mathbf{r},\hat{\mathbf{n}},s)=0,
\end{equation}
with initial condition
\begin{equation}
q_\mathrm{p}(\mathbf{r},\hat{\mathbf{n}},0^+)=\exp[-w_\mathrm{e}(\mathbf{r})].
\end{equation}
From this function, one obtains the partition function of the peptide,
\begin{equation}
{\cal Q}_\mathrm{p}=\int d\mathbf{r}\int d\hat{\mathbf{n}}
\ q_\mathrm{p}(\mathbf{r},\hat{\mathbf{n}},1),
\end{equation}
and, by means of Eqs. (\ref{endden}) and (\ref{coreden}),
its  end and core densities,
\begin{equation}
\label{phie}
\phi_\mathrm{e}(\mathbf{r})=2z_\mathrm{p}\int d\hat{\mathbf{n}}
\ q_\mathrm{p}(\mathbf{r},\hat{\mathbf{n}},1),
\end{equation}
\begin{equation}
\label{phic}
\phi_\mathrm{c}(\mathbf{r})=2z_\mathrm{p}\int_\mathrm{0}^{1/2}ds
\int d\hat{\mathbf{n}}\ q_\mathrm{p}(\mathbf{r},\hat{\mathbf{n}},s)
q_\mathrm{p}(\mathbf{r},-\hat{\mathbf{n}},1-s).
\end{equation}
To summarize: there are five  self-consistent equations to be solved for
the five fields $w_\mathrm{h}(\mathbf{r})$, 
$w_\mathrm{t}(\mathbf{r})$, $w_\mathrm{e}(\mathbf{r})$,
$w_\mathrm{c}(\mathbf{r})$, and $\xi(\mathbf{r})$. 
They are Eqs. (\ref{scf1}) to (\ref{scf5}). 
The fields depend on the four densities $\phi_\mathrm{h}(\mathbf{r})$,
$\phi_\mathrm{t}(\mathbf{r})$, $\phi_\mathrm{e}(\mathbf{r})$, 
and $\phi_\mathrm{c}(\mathbf{r})$, 
which depend, in turn, on these fields. The densities 
are given by Eqs.  
\ref{phih}, \ref{phit}, \ref{phie}, and \ref{phic}.
Once the fields and densities are obtained, the free energy follows 
from Eq. \ref{omega}.

We are interested in the way in which the peptides affect, on the
average, a periodic array of lipid bilayers, that is, a lipid lamellar
phase. We therefore assume that the lamellae are uniform in their plane,
and vary only in the normal direction, specified by the coordinate
$z$. 

In the limit of vanishing peptide density, it is sufficient to
solve for the fields and densities of the pure lipid bilayer, and then
to solve Eq. (\ref{pepend}) for the peptide end-segment distribution
function in the presence of those densities. This is easily done in real
space, since the solution of (\ref{pepend}) is just
\[
q_\mathrm{p}(\mathbf{r}, \hat{\mathbf{n}},s)=\exp \left[-\int_\mathrm{0}^s dt\,
w_\mathrm{e}(\mathbf{r}+t\hat{\mathbf{n}}) [\delta(t)+\delta(t-1)]+
w_\mathrm{c}(\mathbf{r}+t\hat{\mathbf{n}}) \right],
\]
and the fields are provided by Eqs. (\ref{scf3}) and (\ref{scf4}) with
$\phi_\mathrm{e}=\phi_\mathrm{c}=0$.

At non-zero peptide densities, the full set of self-consistent equations
must be solved, and it is more convenient to expand all functions of the
position $\mathbf{r}$ and the director $\hat{\mathbf{n}}$ in terms of a
complete set of basis functions (Wang and Warner 1986, Matsen 1996).
\begin{equation}
g(\mathbf{r},\hat{\mathbf{n}})=
\sum_{l,m=0}^{\infty}g_{l,m}f_{l,m}(z,\cos\theta),
\end{equation}
where $\cos\theta$ is the projection of the unit vector
$\hat{\mathbf{n}}$ onto the $z$ axis. A convenient set is
\begin{eqnarray}
f_{l,0}(z,\cos\theta)&=&(2l+1)^{1/2}P_\mathrm{l}(\cos\theta),\nonumber \\
f_{l,m}(z,\cos\theta)&=&(4l+2)^{1/2}\cos\left({2\pi m z\over D}\right)
P_l(\cos\theta)\qquad {\rm even}\ m,\nonumber \\
f_{l,m}(z,\cos\theta)&=&(4l+2)^{1/2}\sin\left({2\pi m z\over D}\right)
P_l(\cos\theta)\qquad {\rm odd}\ m,
\end{eqnarray}
where $P_l$ is the $l$'th Legendre polynomial, and $D$ is the period of
the lamellae. The latter is determined by minimization of the free
energy with respect to it.  Details of the procedure for solving the
self-consistent equations in this basis can be found elsewhere (Matsen
1996). Of importance here is that the expansion into the infinite set of
basis functions must be truncated to an expansion in a finite number of
such functions. We have utilized $30$ values of $l$ and $10$ values of
$m$, or a total of $300$ functions. The errors in the free energy
brought about by this truncation are less than $0.1\%$.

The parameters we have chosen for our calculations are as follows. The
lipid is characterized by the volume of the headgroup,
$\nu_\mathrm{h}=370$\AA$^3$, and the volume of the tails,
$2N\nu_\mathrm{t}=999$\AA$^3$. For comparison, the volume of the
phosphatidylcholine headgroup is 337\AA$^3$ and that of two tails with
seventeen carbons and one double bond each is 985\AA$^3$ (Armen at al.,
1998). The radius of gyration of the tails was taken to be
$R_\mathrm{g}=12.3$\AA\ which was found to be reasonable in a previous
study (Li and Schick 2000). The peptide is characterized by the volume
of the end groups, which we took to be $\nu_\mathrm{e}=555$\AA$^3$ each,
the volume of each of its core units, $\nu_\mathrm{c}=96.2$\AA$^3$, and
the length of each core group $b$ which we took to be the length of each
amino acid in an alpha helix conformation, 1.5\AA. For comparison, the
volume of the two tryptophans and two alanines, which were only a
portion of the end groups used by de Planque et al. (de Planque et al.,
1998), is 460\AA$^3$, while the average of the volumes of the core
alanine and leucine units is 95.5\AA$^3$. The interaction strength
between hydrophilic and hydrophobic elements is such that
$\chi\gamma_\mathrm{t}=20$. We have taken the number of peptide units to
vary from $L=10$ to $40$.

\section{Results}

We first consider the limit in which the density of peptide is
vanishingly small. The volume fraction distribution of the
lipid headgroups and tails in the anhydrous lipid bilayer are shown in
Fig. 1.  The period $D$, in units of the radius of gyration of the lipid
tails, $R_\mathrm{g}$ is $D/R_\mathrm{g}=2.831$. The thickness of the
hydrophobic region, $D_\mathrm{h}$, as defined by those points at
which the volume fraction of the tails
$\gamma_\mathrm{t}\Phi_\mathrm{t}=0.5$, is $D_\mathrm{h}/D=0.757$.

It is of interest in this limit to determine whether the inserted
peptide spans the bilayer and, if so, whether it inserts normal to the
bilayer or at an angle, $\theta$, to it. This is readily determined.  
We compute the probability
distribution (Matsen, 1996) \[\tilde P_\mathrm{L}(z,\cos\theta)\equiv
q_\mathrm{p}(z,\cos\theta,1) / \int dz d(\cos\theta)
q_\mathrm{p}(z,\cos\theta,1)\] of peptide ends which are at an angle $\theta$ with
the bilayer normal, when that end is located at position $z$ within the
bilayer.  The coordinates $z/D=0$ and $1$ correspond to the center of
the tail region, as in Fig.~1.  The probability distribution can be
calculated for peptides of different length $Lb$. It is shown in Fig. 2
for an $L=20$ peptide, for which $Lb/D_\mathrm{h}=1.143$, that is,
somewhat longer than the hydrophobic thickness of the bilayer. One sees
two major peaks, both of which correspond to a peptide which spans the
bilayer.
One corresponds to a peptide inserted almost normally, at an angle such
that  $\cos\theta\approx -0.9$ and with one end at $z/D\approx 0.4$
This peptide
would pass through the tail region at $z/D=0$. Ends of peptides in the
adjacent lamellae near $z/D\approx 0.6$ are characterized by an angle
$\pi-\theta$ so that the value of the cosine is the negative of that of
the first peak. Thus the second peak simply describes the other end of
the rigid, (and periodically repeated), peptide.

In order to illuminate the behavior of this angle of insertion, we 
integrate the probability distribution of Fig. 2 over all spatial positions
for a fixed angle to obtain the probability
distribution $P_L(\cos\theta)$. It is shown in Fig. 3 for several values of the
hydrophobic length of the peptide $L b$ divided by the hydrophobic
thickness of the
bilayer, $D_\mathrm{h}$. From this distribution we obtain the average
angle of peptide insertion and the most probable angle of insertion as a
function of $Lb/D_\mathrm{h}$. These quantities are plotted in Fig. 4 in
dashed and solid lines respectively. We observe that peptides with
hydrophobic lengths $(Lb/D_\mathrm{h})<(L^*b/D_\mathrm{h})=1.07$, 
insert normal to the bilayer. One might have expected this ratio to be
unity, but it must be recalled that our definition of the hydrophobic
thickness of the membrane, $D_h$, in terms of equal head and tail volume
fractions is a somewhat arbitrary one. Peptides which are longer than $L^*b$
insert an a non-zero angle to the membrane normal.
From energetic arguments alone, one might
expect that $\cos\theta\propto 1/L$ for $L\geq L^*$. Indeed in our
calculation, one sees the dependence on $L\cos\theta$ in the second term
of Eq. (\ref{pepend}). However the incompressibility condition depends
only on the volume of the peptide, proportional to $L$,  
not on its orientation. Thus  the behavior
$\cos\theta\propto 1/L$ should only be  a simple approximation to the actual
behavior. That this is indeed so is seen in Fig. 4 where we have plotted
this dependence as a dotted line. One sees that our results
deviate from this simple description. 

Returning to Fig. 2, we also observe 
two smaller peaks at  $\cos\theta=0$  corresponding to a
fraction of peptides {\em which do not traverse the lamellae}, but are
parallel to it. As our model peptide is predominantly, but not completely,
hydrophobic, the non-traversing peptides are found somewhat below the head-tail interface in the
tail region of the bilayer. In order to determine how this amount  of
non-traversing peptides varies with the peptide length, we calculate the
fraction of peptides with an end at an angle $0\leq\cos\theta\leq 0.2$,
\begin{equation}
I\equiv\int_0^{0.2} P_L(\cos\theta)d(\cos\theta).
\end{equation}
This fraction is plotted {\em vs.} $Lb/D_\mathrm{h}$  in the inset of
Fig. 3. We see that the fraction which do not insert across the membrane
increases with the mismatch between peptide and bilayer hydrophobic
thicknesses, $(Lb/D_h)-1$, irrespective of the sign of the mismatch. The
largest fraction of inserted peptides occurs for $Lb/D_h\approx 1.15$.
 
We now consider non-zero peptide densities. We have calculated the
period, $D$, of
the lamellar phase as a function of peptide volume fraction, $x_p$,  for
values of $x_p<0.04$ and find that $D$
varies essentially linearly with it; {\em i.e.}  
\begin{equation}
\label{slope}
D(x_p)/D(0)\simeq 1+Rx_p,
\end{equation} 
where $D(0)$ is the
period of the lamellar phase in the absence of peptide. 
In Fig. 5 we plot the rate of bilayer thickening, $R$, versus the 
relative peptide thickness $Lb/D_h$. We see that the insertion of 
short peptides,
$Lb/D_h<1.17$, causes
the period, and therefore the bilayer thickness, to contract, while the
insertion of long peptides causes it to increase. As the peptides become
longer, the rate of increase becomes smaller. Eventually the
insertion of sufficiently long peptides, $Lb/D_h>1.6$, causes the
bilayer thickness to {\em decrease}.

\section{Discussion}

Our results replicate and illuminate the experimental
results listed in the Introduction. We recall and discuss them in turn.

{\bf The insertion of peptides whose hydrophobic thickness is less than that
of the bilayer causes a reduction in the bilayer thickness.}

Our calculation reproduces this result. The behavior is
clearly due to simple energetic considerations.

{\bf The insertion of peptides whose hydrophobic thickness is greater
than that of the bilayer causes an increase in the bilayer thickness.
As the
peptides are made longer by a certain amount, $\delta d$, 
the membranes continue to thicken, but the
increase in membrane thickness is less than $\delta d$}.

We reproduce these results. As noted earlier, their origin is  not
obvious,
for energetically the peptides would be satisfied to insert at
an angle such that the bilayer would not deviate from its
original thickness at all. The reason for this behavior, therefore, is
clearly entropic. To understand it, we need only recall the reason 
the bilayer thickness takes the value it does in the absence of
peptide. The repulsive interaction between lipid headgroups and tails
tends to crowd the head groups together which, from the constraint of
incompressibility, causes the lipid tails to stretch. This tendency is
opposed by the loss of tail entropy such stretching brings about. The
equilibrium thickness of the bilayer results from a balance of these two
tendencies. Thus in the equilibrium configuration, the lipid tails are
stretched (Fattal and Ben-Shaul, 1993). Peptides, being rigid, do not
stretch, and therefore lose no such entropy if the bilayer thickens,
while they displace lipid tails which restrain the membrane from
thickening. Thus it can be understood that the insertion of longer
peptides causes the width of the bilayer to increase.  Within this
mechanism, however, there is no
reason that a given increase in the length of the
peptide should result in a corresponding increase in the thickness of the
bilayer. The effect of replacing elastic lipids with rigid
peptides predicts that the insertion of a much bulkier rigid object,
such as gramacidin, will have a larger effect in increasing the bilayer
thickness than will a less bulky one, such as a simple peptide. This is
again in accord with experiment (de Planque, 1999).

{\bf An increase in the magnitude of mismatch, whether due to peptides
being too long or too short for the membrane, results in an increase in
the fraction of peptides which fail to incorporate in the membrane.}

Our calculation reproduces this as shown in the inset to Fig. 3. Short
peptides tend not to insert for energetic reasons. Long peptides tend not
to insert even though there is no energetic penalty to do so.  Presumably, they fail to
insert due to the loss of lipid tail entropy, which increases with the
length of the hydrophobic portion of the inserted peptide. 

In addition to reproducing these experimental results which
demonstrates its efficacy,  our model  also
yields predictions. As noted earlier, we found that insertion of
peptides whose hydrophobic length is greater than the hydrophobic
thickness of the membrane causes the membrane thickness to increase. As
the peptide is made longer, the {\em rate of increase} of membrane thickness  
with peptide length, denoted $R$ in Eq. (\ref{slope}),  
decreases 
in accord with experiment (de
Planque,2000). Strikingly, our model predicts that insertion of peptides
which are very long compared to the hydrophobic thickness of the membrane
$(Lb/D_h>1.6$ in Fig. 5) actually causes the membrane thickness to {\em
decrease}. We believe the reason for this  is that a significant
fraction of these long peptides do not traverse the membrane, but lie
parallel to it. Because the peptides have a much larger end group
relative to its core than do the lipids, they create relatively more
free volume for the lipid tails to fill, {\em i.e.} they effectively
increase the area per lipid head group. As the system is incompressible,
this effect tends to make the bilayer thickness decrease. 
We have recently learned  that this unusal thinning of the bilayer on
the addition of relatively long peptides  has been observed 
(Killian, 2001).

Finally, our model makes predictions about the insertion of peptides at
very low densities. Short peptides, whose hydrophobic thickness is less
than the hydrophobic thickness of the bilayer, insert normal to
it. Peptides with a hydrophobic thickness greater than that of the
bilayer insert at an angle to the normal which grows with
the hydrophobic mismatch. We expect the same behavior for proteins which
span the membrane with a single alpha helix.
As we have ignored local effects in the plane of the
membrane, the actual angle of insertion will differ somewhat from that
which we have calculated, but  the qualitative
behavior will not be changed. 
Except for very large mismatches, our theory predicts
the tilt to decrease with increasing peptide concentration. 
Although there is much data on helix tilt in specific systems, there
appears to be no attempt at a systematic correlation of it with
hydrophobic mismatch. Such data would be most interesting.
 
We thank Prof. J. A. Killian for useful comments and providing the
Ph. D. thesis of M. R. R. de Planque.  This work is supported by the NSF
under grant DMR9876864. D.D. acknowledges support from a Spanish
Ministry of Education, Culture and Sports MEC-FPI EX 2000 grant.

\newpage

\begin{center} 
{\bf REFERENCES} 
\end{center} \begin{description} 
\item{}Armen, R., O. Uitto, and S. Feller. 1998. Phospholipid component
volumes: determination and application to bilayer structure
calculations. {\em Biophys. J.} 75:734-744.  
\item{} de Planque,M. R. R. 2000.  
Hydrophobic matching and interfacial anchoring of
transmembrane peptides as determinants for membrane
organization. Ph. D. Thesis, University of Utrecht.  
\item{} de Planque,
M. R. R., D. V. Greathouse, R.  E. Koeppe II, H. Sch{\"a}ffer, D. Marsh,
and J. A. Killian, 1998. Influence of lipid/peptide hydrophobic mismatch
on the thickness of diacylphosphatidylcholine bilayers. A $^2H$ NMR and
ESR study using designed transmembrane $\alpha$-helical peptides and
gramacidin A. {\em Biochemistry} 37:9333-9345.  
\item{} de Planque,
M. R. R. , J. A. W. Kruijtzer, R. M. J. Liskamp, D. Maarsh,
D. V. Greathouse, R. E. Koeppe II, B. de Kruijff, and J. A. Killian,
Different membrane anchoring positions of tryptophan and lysine in
synthetic transmembrane $\alpha$-helical peptides. {\em J. Biol. Chem.}
274:20839-20846.  
\item{}Dumas, F., M. C. Lebrun, and
J.-F. Tocanne. 1999 Is the protein/lipid hydrophobic matching principle
relevant to membrane organization and functions? {\em FEBS Letters}
458:271-277.  
\item{}
Edelman, J. 1993. Quadratic Minimization of Predictors for Protein Secondary
Structure: Application to Transmembrane $\alpha$-Helices. {\it J. Mol. Biol.}
232:165-191. 
\item{} Fattal, D. R. and A. Ben-Shaul. 1993. A molecular
model for lipid-protein interaction in membranes: the role of
hydrophobic mismatch. {\em Biophys. J.} 65:1795-1809.  
\item{}Froud,
R.J., C. R. A. Earl, J. M. East, and A. G. Lee. 1986. Effects of lipid
fatty acyl chain structure on the activity of the (Ca2+-Mg2+)-ATPase.
{\em Biochim. Biophys. Acta} 860:416-421.  
\item{} Johannsson, A.,
G. A. Smith, and J. C. Metcalfe. 1981. The effect of bilayer thickness
on the activity of (Na+ + K+)-ATPase. {\em Biochim. Biophys. Acta}
641:416-421.  
\item{} Killian, J.A., I. Salemink, M.R.R. de Planque, G. Lindblom,
R.E. Koeppe II., and D.V. Greathouse. 1996 Induction of nonbilayer
structures in diacylphosphatidylcholine model membranes by transmembrane
$\alpha$-helical peptides: Importance of hydrophobic mismatch and
proposed role of tryptophans. {\em Biochemistry} 35:1037-1045.
\item{} Killian, J.A. 1998 Hydrophobic mismatch between proteins and
lipids in membranes. {\em Biochim. Biophys. Acta} 1376:401-416.  
\item{} Killian, J. A. 2001., private communication.
\item{} Lehtonen, J. Y. A. and P. K. J. Kinnunen. 1997. Evidence for
phospholipid microdomain formation in liquid crystalline liposomes
reconstituted with Esherichia coil lactose permease. {\em Biophys. J.}
72:1247-1257.  
\item{} Li, X.-J., and M. Schick. 2000. Theory of Lipid
Polymorphism: Application to Phosphatidylethanolamine and
Phosphatidylserine. {\em Biophys. J} 78:34-46.  
\item{} Marsh,
D. 1995. Lipid-protein interactions and heterogeneous lipid distribution
in membranes. {\em Mol. Membr. Biol.} 12:59-64.  
\item{} Matsen,
M. W. 1996. Melts of semiflexible diblock copolymer, {\em
J. Chem. Phys.} 104:7758-7764.  
\item{} May, S. 2000. Theories of
structural perturbations of lipid bilayers. {\em Curr. Opinion in
Coll. and Interface Sci.} 5:244-249.  
\item{} May, S. and
A. Ben-Shaul. 1999.  Molecular theory of lipid-protein interaction and
the $L_{\alpha}-H_{II}$ transition. {\em Biophys. J. } 76:751-767.
\item{} Mouritsen, O. G., and M. Bloom. 1984. Matress model of
lipid-protein interactions in membranes.  {\em Biophys. J.} 46:141-153.
\item{}Pelham, H.R. and S. Munro. 1993. Sorting of membrane proteins in
the secretory pathway. {\em Cell} 75:603-605 
\item{}Piknova,
B. E. Perochon, and J. F. Tocanne. 1993. Hydrophobic mismatch and
long/range protein lipid interactions in
bacteriorhodopsin/phosphatidylcholine vesicles. {\em Eur. J. Biochem.}
218:385-396.  
\item{} Ren, J., S. Lew, Z. Wang, and
E. London. 1997. Transmembrane orientation of hydrophobic
$\alpha$-helices is regulated both by the relationship of helix loength
to bilayer thickness and by the cholesterol concentration. {\em
Biochemistry} 36:10213-10220.  
\item{} Ren, J., S. Lew, J. Wang, and
E. London. 1999. Control of the transmembrane orientation and
interhelical interactions within membranes by hydrophobic helix
length. {\em Biochemistry} 38:5905-5912.  
\item{}
Rost, B., R. Casadio, P. Fariselli, and C. Sander. 1995. Transmembrane
helices predicted at 95\% accuracy. {\it Protein Sci.} 4:521-533.
\item{} van der Wei, P. C. A.,
T. Pott, S. Morein, D. V. Greathouse, R. E. Koeppe II, and
J. A. Killian. 2000. Tryptophan-anchored transmembrane peptides promote
formation of nonlamellar phases in phosphatidyletanolamine model
membranes in a mismatch-dependent manner. {\em Biochemistry}
39:3124-3133.  
\item{} Wang, X. and M. Warner. 1986. Theory of nematic
backbone polymer phases and conformations. {\em J. Phys. A:Math. Gen.}
19:2215-2227.  
\end{description} 
\newpage 
\begin{description}
\item[Figure 1] Volume fraction distribution of the lipid headgroups and
tails in the lamellar phase plotted \textit{vs.} coordinate $z$
perpendicular to the lamellae. The wavelength of the phase is $D$.
$z/D=0.5$ corresponds to the center of the head region, and $z/D=0$ and
$1$, to the center of the tail region.  The interaction strength between
hydrophobic and hydrophilic entities is such that
$\chi\gamma_\mathrm{t}=20$.  
\item[Figure 2] Normalized probability distribution,
${\tilde P}_\mathrm{L}(z,\cos\theta)$ of the orientation, $\cos\theta$,
and location, $z$, of the peptide ends for a $L=20$ peptide, of length
$Lb/D_\mathrm{h}=1.143$, with $D_\mathrm{h}$ the hydrophobic thickness
of the lamellae. The center of the lipid tail region is at $z/D=0$ and
$1$. Contour values are given in the legend.  
\item[Figure 3]
Normalized probability distribution $P_L(\cos\theta)$ 
of the orientation of the
peptide ends  \textit{vs.} $\cos\theta$ 
for various relative peptide lengths $Lb/D_\mathrm{h}$, as noted on each
curve. The lengths 
correspond to $L=16$ to $L=26$ in increments of $2$, plus
two extreme cases, $L=10$ and $L=40$. The inset shows the fraction, $I$, of
peptides which do not insert into the bilayer as a function of 
 $Lb/D_\mathrm{h}$.  
\item[Figure 4] Full line:
location of the peak in the probability distribution function
$P(\cos\theta)$ which occurs at non-zero values of $\cos\theta$; dashed
line: mean value of $\cos\theta$; dotted line: simple $1/L$
behavior. Curves are given as functions of peptide length in units of
$D_\mathrm{h}$ (lower $x$ scale) and $L$ (upper $x$ scale).
\item[Figure 5] Rate, $R$, of increase of relative 
lamellar period with volume fraction, $x_p$, 
of peptide {\em vs.} relative hydrophobic length of peptide.  
\end{description}
\newpage
\centering
\includegraphics[width=\linewidth]{fig1}
\vskip0.4in{\Large Fig. 1}
\newpage
{
\psfrag{costheta}[][][2]{$\cos\theta$}
\psfrag{zoverD}[][][2]{$z/D$}
\centering
\resizebox{\linewidth}{!}{%
\includegraphics{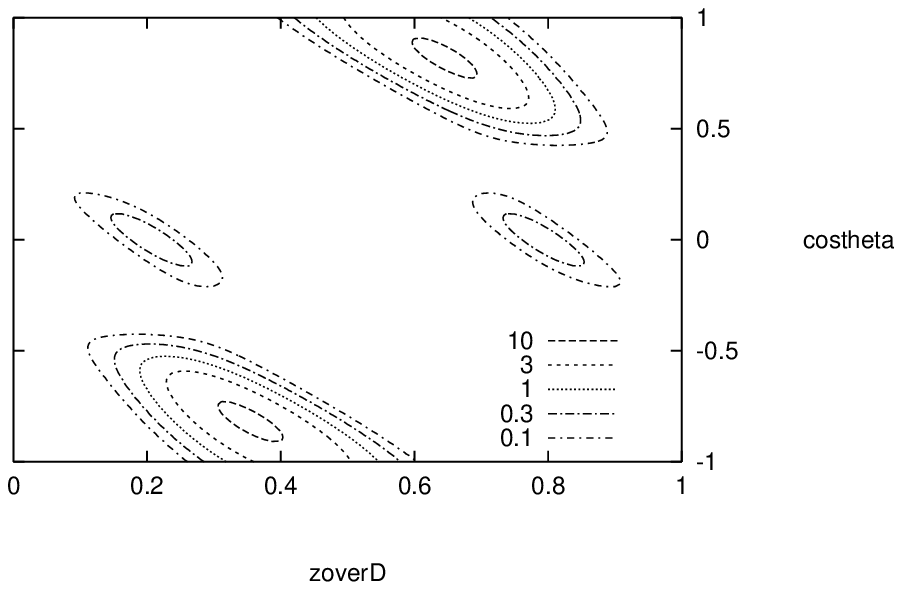}}
\vskip0.4in{\Large Fig. 2}
}
\newpage
\centering
\includegraphics[width=\linewidth]{fig3}
\vskip0.4in{\Large Fig. 3}
\newpage
\centering
\includegraphics[width=\linewidth]{fig4}
\vskip0.4in{\Large Fig. 4}
\newpage
\centering
\includegraphics[width=\linewidth]{fig5}
\vskip0.4in{\Large Fig. 5}
\end{document}